
%
\voffset=-50pt
\magnification=1200
\baselineskip=15pt
\overfullrule=0pt
\hsize=17.2truecm
\vsize=25.5truecm
\nopagenumbers
\font\ninerm=cmr9
\font\eightrm=cmr8
\centerline {\bf STELLAR CONTENT AND STAR FORMATION HISTORIES}
\centerline {\bf OF IRREGULAR GALAXIES}
\medskip\centerline {Monica Tosi}
\par\centerline {Osservatorio Astronomico, Via Zamboni 33, Bologna, Italy}
\parindent=1.0truecm
\vskip 10.5truemm\noindent
{\bf 1. Introduction}
\medskip\noindent
Irregular galaxies are characterized by three major observational features:
their stellar populations are mostly young, their metallicity is low
and their gas content is large. Exactly twenty years ago, Searle, Sargent
and Bagnuolo (1973) published a famous paper where they suggested that to
explain these features either these galaxies have started to form stars only
recently or they have experienced a very discontinuous star
formation (SF) activity. This is because the extrapolation throughout a galaxy
lifetime of $\sim$15 Gyr of the current star formation rate (SFR) inferred from
the observed young stellar populations implies much larger metal contents
and much smaller gas contents than observed. The successive detection of
an underlying old stellar population (e.g. Hunter and Gallagher 1985, Salzer
et al. 1991) has however ruled out in most cases the first hypothesis
and given more weight to that of discontinuous SF processes.
\par
Blue Compact galaxies (BCGs), a special sub-type of irregulars, have indeed
been suggested to experience short and intense episodes of global SF (e.g.
Hodge 1989) and all the chemical evolution models able to reproduce their
observed properties (e.g. Matteucci and Tosi 1986, Pilyugin 1993) require
only a few of these SF bursts during their life. For objects like I Zw 18
(e.g. Kunth et al. 1993) and the Arecibo HI cloud (Salzer et al. 1991) a
maximum of two bursts have been suggested. On the other hand, Hunter
and Gallagher (1985, 1986) have argued that giant irregulars (and possibly
all the Magellanic irregulars, either giant or dwarf) seem to support a
continuous, and even constant, SF.
\par
To analyse the SF regime in irregular galaxies and try to better understand
the difference (if any) among the various morphological subtypes, we are
studying these galaxies from two points of view. On the one hand, G.Marconi,
F.Matteucci and myself are computing detailed models of galactic chemical
evolution and, on the other hand, L.Greggio, G.Marconi and myself have
developed
a method to derive the star formation history in a local sample of irregulars
from their empirical Colour-Magnitude Diagrams (CMDs) and Luminosity Functions
(LFs). The latter study can provide fairly stringent constraints on the
chemical
evolution models; we thus hope that the combination of the two approachs will
lead to a better understanding of the evolution of irregular galaxies.
\bigskip\noindent
{\bf 2. Chemical evolution models}
\medskip\noindent
We have recently computed updated models for the chemical evolution of BCGs
(Marconi et al. 1994) for which reliable data are available on the element
abundances. We have also started modeling the evolution of the dwarf
(Magellanic) irregulars examined in the next Section, but this work is still in
a very preliminary phase. In the following, I will therefore describe only the
model results concerning BCGs.
\par
These models are based on the Self Propagating Star Formation theory proposed
by Gerola et al. (1980), assume both galactic accretion and loss of gas, adopt
various choices for the Initial Mass Function (IMF), take the actual stellar
lifetimes into account (i.e. do not assume instantaneous recycling) and
follow the evolution of several chemical elements. A detailed description
of the model assumptions and prescriptions is given by Marconi et al. (1994).
\topinsert\vskip 6.5truecm\noindent
{\eightrm Fig.1. a) Helium vs oxygen distribution derived from Pagel's et al.
(1992) data (full circles) and from some of our chemical evolution models
(lines and dots). b) Same for the nitrogen/oxygen vs oxygen distribution.
The average observational error bar is shown in the bottom
right corner of both panels.}
\phantom { }\endinsert
\par
The model predictions have been compared with the empirical abundances of
helium, nitrogen and oxygen derived by Pagel et al. (1992) from spectral
observations of a large sample of HII regions in BCGs and Magellanic
irregulars. Fig.1a shows the He/H versus O/H abundance ratios by number derived
from the models and the data and Fig.1b shows the corresponding N/O versus O/H
abundance ratios in the traditional logarithmic units.
\par
Models assuming no loss of gas from the system allow the galaxy to retain all
the metals synthesized by their stars and predict definitely too much oxygen
in comparison with the data, as shown for instance by the rightmost solid line
of Fig.1b. In Fig.1a their resulting abundances lie out of the box. If a
galactic wind triggered by SNe II and ejecting all the elements in the same
proportions is assumed (and we call it {\it normal} wind),
the resulting oxygen abundances are lower and consistent
with the data, but the helium and nitrogen abundances are reduced as well. As
examplified by the two solid lines of Fig.1a, in this case helium turns out to
be
always underabundant, whatever the assumptions on the other free parameters.
If, instead, we assume a {\it differential} wind, ejecting in much larger
proportions the elements like oxygen synthesized in massive stars, i.e. in the
stars whose explosions induce the wind, we can
properly reduce the O abundance without affecting too much either helium or
nitrogen and predict abundances consistent with the corresponding constraints.
\par
To test which combinations of the free parameters lead to predictions
consistent
with the empirical constraints, we have run a large number of models varying
the slope, the upper and the lower mass limit of the IMF, varying the number,
the starting epochs and the duration of the
bursts and varying also the main nucleosynthesis parameters. To summarize the
obtained results, we can reach a satisfactory agreement with the observational
distributions only if:
\par
\leftskip 0.5truecm
\parindent -0.5truecm
\par
$-$ The SF proceeds in short ($\sim 10^6$yr), intense (SFR$\gg$SFR$_{\odot}$)
and well separated ($\sim 10^9$yr) bursts.
\par
$-$ In most cases no more than 7 $-$ 10 bursts have taken place throughout the
galaxy lifetime. In a few cases only one or two. The last burst is currently
active.
\par
$-$ The IMF is close to Salpeter's slope $\alpha=-2.35$ with no need for upper
and/or lower mass limits very different from those suggested for the solar
neighbourhood stars.
\par
$-$ Differential galactic winds are present.
\par
\leftskip 0truecm
\parindent 1truecm
\bigskip\noindent
{\bf 3. SF histories in nearby dwarf irregulars}
\medskip\noindent
Since the most direct information on the stellar populations and relative
histories in any system can be derived from their CMD, we have undertaken
a project for accurately studying the CMDs and LFs of nearby, well resolved,
irregulars. The method (Tosi et al. 1991) proceeds on two tracks: a) deep and
accurate CCD photometry of several galaxies is obtained and the corresponding
CMDs and LFs derived; b) a numerical code for Montecarlo simulations of CMDs
has
been developed and applied to the empirical data. We will see in the following
that the comparison of the observational data with the corresponding
theoretical predictions provides several constraints on the SF history and on
the IMF of the analysed objects.
\par
The simulation code is based on complete sets of homogeneous stellar evolution
tracks of various metallicities. It is very important to adopt homogeneous
sets of models to avoid the appearance in the CMDs of spurious features due
to the interpolation between inconsistent tracks. Unfortunately, there are
only a few complete sets of homogeneous models already available and, so far,
the only ones appropriate for the galaxies of our sample are those computed
by the Padova group (namely those with Z=0.001 and large overshooting from
convective cores presented by Greggio 1984 and Bertelli et al. 1986,
hereinafter BBC86; those with Z=0.008 and moderate overshooting published by
Alongi et al. 1993, hereinafter BBC93; and those with Z=0.008 and without
overshooting presented in the same paper, hereinafter BBC93CL).
\par
The code provides synthetic CMDs with the same number of stars as observed
in each galactic field and takes into account all the theoretical parameters
affecting the evolution of a stellar population (age, metallicity, IMF and SFR)
and all the observational uncertainties (photometric errors, stellar blend and
incompleteness factors). Therefore a model can be considered satisfactory
only if it reproduces all the features of the corresponding observational CMD
and LFs. For this reason, this approach represents a significant improvement
over the classical isochrone fitting method.
\par
The galaxies of our sample are listed in Table 1, where their coordinates,
derived distance modulus, number of observed fields, number of stars detected
in both the B and V bands (and also in the R band, when observed) and
number of selected stars with formal photometric error $\sigma <$ 0.1 mag
are also given. All the galaxies were selected from the DDO
Catalogue (van den Bergh 1959) and were supposed to belong to the Local
Group. A posteriori, we have however found that DDO 210 lies
beyond the edge of the Group. For each galaxy we have observed at least
two internal fields to check if these small size objects can be taken as
homogeneous bodies and one external field for decontamination purposes.
\medskip
\centerline {{\bf Table 1.} Local Group irregulars in our program}
{\eightrm
\baselineskip=10pt
\tabskip=1em plus.5em minus.5em
\def\mrule{\noalign{\vskip6pt\hrule\vskip6pt}}
\mrule
\noindent
\halign to
\hsize{#\hfil&#\hfil&#&#\hfil&#&\hfil#&#&
\hfil#&#&\hfil#\hfil&#&\hfil#\hfil&#&\hfil#\hfil&#&\hfil#\hfil\cr
{}~~~~~~~~~~Name&~~~R.A.&&~~~~~DEC&&l~~~&&b~~~&&(m$-$M)$_o$&&NF&&detected&&selected\cr
\mrule
DDO ~~70 (Sextans B)&09 57 23&&+05 34 07&&233.2&&+43.8&&25.6 &&2&&2434&&1093\cr
DDO 209 (NGC 6822) &19 42 07&&$-$14 55 01&&~25.3&&$-$18.4&&23.5
&&3&&2031&&1772\cr
DDO 210            &20 44 08&&$-$13 02 00&&~34.1&&$-$31.4&&28?&&2&&1247&&
583\cr
DDO 221 (WLM)      &23 59 23&&$-$15 44 06&&~75.9&&$-$73.6&&25.0 &&3&&2513&&
$-$\cr
DDO 236 (NGC 3109) &10 00 48&&$-$25 55 00&&262.1&&+23.1&&25.7
&&3&&6430&&2605\cr
}
\mrule }
\medskip\par
All the observations have been made at the ESO/MPI 2.2 m telescope in La
Silla (Chile) and the data have been reduced with Daophot (Stetson 1987).
We have generally detected stars down to B, V and R magnitudes $\simeq$ 26.5,
but for sake of accuracy we retain only those with $\sigma_{Daophot} <$
0.1 mag, which in practice are all brighter than 24. Given the distances
of the sample galaxies, this limiting magnitude roughly corresponds to
2 M$_{\odot}$ stars; and since their lifetime is $\sim$1 Gyr this is the
backtime limit to our investigation on the star formation history.
More detailed descriptions of the data acquisition and reduction, as well as
the results of their theoretical interpretation,  can be
found in Tosi et al. (1991) for Sextans B, in Marconi et al. (1994) for
NGC 6822, in Greggio et al. (1993) for DDO~210 and NGC 3109 and in
Ferraro et al. (1989) for WLM.
\par
There are two results common to all the galaxies of our sample that I wish
to emphasize:
\par\noindent
1) Special care must be devoted to evaluate in each frame the incompleteness
factors due to both background and blend of stellar images, because they
sensibly affect the stellar distribution in the CMD and have therefore
important effects on their interpretation;
\par\noindent
2) A safe criterion for the selection of MS stars is required
to avoid erroneous derivation of the IMF.
\par\noindent
The above conclusions can be better explained examining in more detail one
of the cases. Here I will focus on Sextans B which has been the first studied
galaxy of our sample.
\par
The two observed fields of Sextans B (which cover the entire galaxy) have
properties very similar to each other and panel a) of Fig.2 shows the CMD
derived for the more populated region. The large dispersion of the stellar
distribution, the  bright blue plume and the number of red stars significantly
lower than the number of blue stars are all typical of the CMDs of Magellanic
irregulars.
\topinsert\vskip 5truecm\noindent
{\eightrm Fig.2. a) Observational CMD of Region 2 of Sextans B; b) synthetic
CMD for the same region; c) same synthetic CMD as in b) but without taking
stellar blend into account; d) synthetic CMD as in b) with 10 times longer
quiescent phase.}
\phantom { }\endinsert
\par
The other panels of the figure show
some of the corresponding synthetic CMDs based on the BBC86 stellar models.
In panel b) one of our preferred models for Sextans B is presented. The
only major flaw of this synthetic CMD is the large number of bright red
supergiants which are not present in the observational CMD. This is an
inconsistency of the adopted set of tracks (the only ones available and
compatible with the data three years ago) which has been
later overcome in the new sets computed in Padova. The model of panel b)
assumes
an IMF with slope $\alpha=-$2.6 and two SF episodes in the last 1 Gyr,
interrupted by a short quiescent phase.
The SF activity in both episodes is rather modest with a SFR two to three
orders of magnitude lower than in BCGs. The quiescent interval between the
two episodes lasts only 20 Myr; to avoid
the presence of too bright, non observed, stars no SF activity is assumed
in the last 3 Myr. This regime is quite different from that of BCGs and we
prefer to call it {\it gasping} rather than {\it bursting}. The LF derived
from the main sequence (MS) stars of this model is in agreement with the
corresponding empirical MS LF, as well as the global LF.
\par
The synthetic diagram of Fig.2b takes into account all the photometric errors,
including the stellar blend of objects which happen to be too close
in the projected plane of a CCD frame to be distinguishable from each other.
This blend leads to assign spurious magnitudes and colours to the detected
objects and is responsible for a large fraction of the spread in the stellar
distribution in the CMD. If stellar blend is not included the same synthetic
diagram becomes that of Fig.2c where the observational spread is now due only
to the formal photometric error $\sigma$. The stellar sequences appear much
tighter and the various evolutionary phases are much more separated from each
other. In particular, the blue plume is now splitted into two parts: objects
at the left of the vertical gap are main sequence stars, objects at
its right are evolved stars at the hot edge of the blue loop evolutionary
phase. This implies that the observational blue plume is populated by
both MS and evolved stars. This finding represents a serious warning for
people who derive the LF of the blue plume and adopt it as MS LF to infer
the IMF of the region. For a correct evaluation of the IMF a reliable
criterion is needed to select the actual MS stars from the whole blue
plume population.
\par
The diagram of Fig.2d is equivalent to that of Fig.2b, with the recent SF
episode exactly equal to that of Fig.2b but with a shorter duration of the
older episode, so that the quiescent phase now lasts 250 Myr. Such a long
interval leads to the large, almost horizontal, gap in panel d), which
separates
the stars born in the two different episodes. It is apparent that the gap
is not consistent with the observed CMD of panel a). With similar checks, we
can place an upper limit of $\sim$ 100 Myr to the quiescent phase in Sextans B
and it is interesting to compare this duration with that of the quiescent
phases in BCGs which are at least ten times longer.
\topinsert\vskip 17truecm\noindent
{\eightrm Fig.3. Observational CMDs of the observed fields in NGC 6822 (left),
NGC 3109 (center) and DDO 210 (right).}
\phantom { }\endinsert
The left panel of Fig.3
shows the CMDs of the stars detected with $\sigma_{Daophot} <$ 0.1 mag
in B and V in the three observed fields of NGC 6822 and in its external field.
The usefulness of the latter is demonstrated by the presence in the CMDs
of all the internal regions of a vertical sequence of objects around
B-V $\simeq$ 0.9 corresponding exactly to the location in the CMD of the
foreground stars detected in the external field. The blue edge of the blue
plume of Region A seems slightly redder than that of Region C, thus suggesting
either a larger reddening or a larger metallicity or a combination of both.
\topinsert\vskip 5truecm\noindent
{\eightrm Fig.4. Synthetic CMDs for Region A of NGC 6822: a) best model based
on the BBC93CL tracks, b) best model based on the BBC93 tracks and c) best
model based on the BBC86 tracks.}
\phantom { }\endinsert
\par
Fig.4 shows the CMDs derived by the three best models for Region A based on
three sets of stellar evolution tracks: the CMD in panel a) is based on
the BBC93CL tracks with no overshooting and assumes an IMF slightly flatter
($\alpha$=-2.2) than Salpeter's and two {\it gasps} of SF in the last 1 Gyr
stopped 4 Myr ago; the CMD in panel b) is based on the BBC93 tracks with
moderate overshooting and assumes Salpeter's IMF ($\alpha$=-2.35) and two
{\it gasps} of SF stopped 3 Myr ago; the CMD of panel c) is based on the BBC86
tracks with large overshooting and assumes a flatter IMF ($\alpha$=-2.0) and
a constant SF stopped 2 Myr ago.
Fig.5 shows the LFs derived for all the stars (right panel) and
for MS stars (left panel) and compared with the corresponding empirical
data: all the three models reproduce pretty well the observational functions.
The individual epochs of the SF episodes and the corresponding rates slightly
differ from each other, as well as the derived IMF slopes, and this is due
to the different characteristics of the adopted tracks. It is remarkable,
however, that despite the uncertainties related to the choice of the stellar
evolution models all the three best simulations indicate a moderately flat
IMF, a modest SFR, possibly interrupted for a very short time around 100
Myr ago and no activity in the last few Myrs.
\bottominsert\vskip 5.55truecm\noindent
{\eightrm Fig.5. Luminosity functions for Region A of NGC 6822. The dots
correspond to the empirical values, the dotted lines to the model of Fig.4a,
the dashed lines to the model of Fig.4b and the solid lines to the model
of Fig.4c.}
\phantom { }\endinsert
\medskip\par
The central panel of Fig.3
shows the CMDs of the stars detected with $\sigma_{Daophot} <$ 0.1 mag
in B, V and R in the three observed fields of NGC 3109 and in its external
field. The blue edge of the blue plume of Region C is significantly redder
than that of Region A, thus suggesting a larger reddening and/or a larger
metallicity. It is worth noticing that the two regions differ also in the
red supergiant populations: despite the lower total number of stars (933
compared with 1019 in Region A), the CMD of Region C contains definitely
more stars with V$\leq$21 and B-V$\geq$1.0.
\par
The right panel of Fig.3
shows the CMDs of the stars detected with $\sigma_{Daophot} <$ 0.1 mag
in B and V in the two observed fields of DDO 210 and in its external field.
In this case the external field is particularly important because the system
is highly contaminated both by foreground stars and by background galaxies.
The objects brighter than V=21 are presumably
all foreground stars and those fainter than V=22 are mostly members of
DDO 210 (except for a few background galaxies). In between there is a mixture
of members (mostly blue) and non members (mostly red). Within the uncertainties
the two observed regions of this galaxy seem to contain the same stellar
populations.
\topinsert\vskip 5truecm\noindent
{\eightrm Fig.6. Observational CMDs of Regions 1 and 2 of WLM. Region 2
corresponds to two adjacent CCD fields containing the same stellar populations.
}
\phantom { }\endinsert
\par
Finally, Fig.6 shows our most striking example of a small system ($\sim$
2 kpc across) with very different populations in different regions. Despite
the much larger number of detected objects, Region 2 of WLM contains no blue
stars brighter than V$\simeq$20, whereas the blue plume of its Region 1
extends up to V=18, two magnitudes brighter ! This implies that the SF activity
has been very recently efficient in Region 1, but has stopped between 50 and
100 Myr ago in Region 2.
\bigskip\centerline {{\bf Table 2.} Results from simulations.}
\par
{\eightrm
\baselineskip=7pt
\tabskip=1em plus.4em minus.4em
\def\mrule{\noalign{\vskip6pt\hrule\vskip6pt}}
\mrule
\noindent
\halign to
\hsize{#\hfil&\hfil#\hfil&#&#&\hfil#\hfil&#&\hfil#\hfil&\hfil#\hfil\cr
Name&Homogeneity&&&IMF&&SF type&SFR\cr
\mrule
Sextans B &yes &&     & 2.6 && gasps stopped & SFR$_{young}\simeq$0.5
SFR$_{old}$ \cr
\mrule
          &    &&Reg.A& 2.2 && gasps stopped  & SFR$_{young}\simeq$0.5
SFR$_{old}$ \cr
          &    &&     & 2.35&& const stopped  & SFR$_{young}\simeq$1.0
SFR$_{old}$ \cr
NGC 6822  &no  &&$---$&$---$&&$-------$&$--------------$ \cr
          &    &&Reg.C& 2.6 && gasps stopped   & SFR$_{young}\simeq$0.4
SFR$_{old}$ \cr
          &    &&     & 2.35&& const stopped? & SFR$_{young}\simeq$0.2
SFR$_{old}$ \cr
\mrule
DDO 210   &yes &&     & 2.35&& gasps stopped   & SFR$_{young}\simeq$0.1
SFR$_{old}$ \cr
\mrule
          &    &&Reg.1& 2.35&& gasps stopped   & SFR$_{young}\simeq$2.0
SFR$_{old}$ \cr
WLM       &no  &&$---$&$---$&&$-------$&$--------------$ \cr
          &    &&Reg.2& 2.35?&& stopped 5-10 Myr ago  &\cr
\mrule
          &    &&Reg.A& 1.2 && gasps stopped   & SFR$_{young}\simeq$0.3
SFR$_{old}$ \cr
          &    &&     & 2.35&& gasps stopped   & SFR$_{young}\simeq$2.0
SFR$_{old}$ \cr
NGC 3109  &no  &&$---$&$---$&&$-------$&$--------------$ \cr
          &    &&Reg.C& 2.35&& gasps stopped   & SFR$_{young}\simeq$3.0
SFR$_{old}$ \cr
}
\mrule
}
\medskip\par
For all the dwarfs of our sample we have performed a large number of
simulations
of the kind described for Sextans B and NGC 6822. Their detailed results can
be found in the papers quoted above. Here I only summarize the overall
results (see also Table 2):
\par
\leftskip 0.5truecm
\parindent -0.5truecm
\par
$-$ Despite their small sizes, only two out of five galaxies can be considered
as single homogeneous bodies; the other three contain different populations
in different regions.
\par
$-$ The IMFs in better agreement with the observed populations do not deviate
much from Salpeter's slope $\alpha=-2.35$.
\par
$-$ In most cases the appropriate SF regime is a {\it gasping} activity (i.e.
long episodes with modest SFR separated by short quiescent intervals).
\par
$-$ The SFR inferred for the two episodes of the last 1 Gyr are generally
different from each other.
\par
$-$ In most cases the SF processes must have been inactive in the last few Myr.
For NGC 6822 and NGC 3109 there is however a selection effect due to our
choice of observing fields not containing HII regions to avoid reduction
problems. In these two galaxies several HII regions are present, indicating
that somewhere they do currently form stars.
\par
\leftskip 0truecm
\parindent 1truecm
\par
Except for the IMF, the above results are quite different from those derived
for BCGs and described in the previous section. Does this suggest that
BCGs and Magellanic irregulars are systems with completely different histories
(i.e. formation and/or evolution) ? Or is it a size effect as predicted by
the Stochastic Self Propagating theory of SF proposed by Gerola et al. (1980)
and should we interpret in terms of increasing galactic dimensions the sequence
from bursting SF in BCGs to gasping SF in dwarf irregulars to
continuous/constant SF in giant irregulars ? We hope that the computation of
chemical evolution models for these well studied dwarfs will provide further
information on their previous history and give some hints to answer these
questions.

\bigskip
I wish to thank Laura Greggio, Francesca Matteucci and Gianni Marconi
with whom most of these studies have been done.
\medskip\noindent
{\ninerm
{\bf References}
\medskip\noindent
Alongi, M., Bertelli, G., Bressan, A., Chiosi, C., Fagotto, F., Greggio, L.,
 Nasi, E., 1993, {\it A.A.S.S.}, \par {\bf 97}, 851
\par\noindent
Bertelli G., Bressan A., Chiosi C., Angerer K. 1986, {\it A.A.S.S.}
{\bf 66}, 191.
\par\noindent
Ferraro, F.R., Fusi Pecci, F., Tosi, M., Buonanno, R. 1989, {\it M.N.R.A.S.}
{\bf 241}, 433
\par\noindent
Gerola, H., Seiden, P.E. and Schulman, L.S., 1980, {\it Ap.J.} {\bf 242}, 517
\par\noindent
Greggio L., 1984 in {\it Observational Tests of the Stellar Evolution Theory},
A.Maeder and \par A.Renzini eds (Dordrecht:Reidel), p. 329.
\par\noindent
Greggio, L., Marconi, G., Tosi, M., Focardi, P., 1993, {\it A. J.}
 {\bf 105}, 894
\par\noindent
Hodge, P., 1989, {\it A.R.A.A.} {\bf 27}, 139
\par\noindent
Hunter, D.A. and Gallagher, J.S.III, 1985, {\it Ap.J. S.S.} {\bf 58},533
\par\noindent
Hunter, D.A. and Gallagher, J.S.III, 1986, {\it P.A.S.P.} {\bf 98}, 5
\par\noindent
Kunth, D., Lequeux, J., Sargent, W.L.W., Viallefond, F. 1993,
{\it A.A.} in press
\par\noindent
Marconi, G., Matteucci, F., Tosi, M. 1994, {\it M.N.R.A.S.}, submitted
\par\noindent
Marconi, G., Tosi, M., Greggio, L., Focardi, P. 1994, {\it A.J.} submitted
\par\noindent
Matteucci, F. $\&$ Tosi, M. 1985, {\it M.N.R.A.S.} {\bf 217}, 391
\par\noindent
Pagel, B.E.J., Simonson, E.A., Terlevich, R.J., Edmunds, M.G., 1992,
{\it M.N.R.A.S.} {\bf 255}, 325
\par\noindent
Pilyugin , L.S. 1993 {\it A.A.} in press
\par\noindent
Salzer, J.J., Di Serego-Alighieri, S., Matteucci, F.,
 Giovanelli, R., Haynes, M.P., 1991, {\it Ap. J.} {\bf 101}, \par 1258
\par\noindent
Searle, L., Sargent, W.L.W., Bagnuolo, W. 1973, {\it Ap.J.} {\bf 179}, 427
\par\noindent
Stetson, P.B. 1987, {\it P.A.S.P.} {\bf 99}, 191
\par\noindent
Tosi, M., Greggio, L., Marconi, G., Focardi, P. 1991, {\it A.J.} {\bf 102},
951
\par\noindent
van den Bergh, S. 1959, {\it A Catalogue of Dwarf Galaxies} (DDO Publications,
Toronto)
}
\bye